\documentclass[prb,twocolumn,showpacs,amsmath,amssymb]{revtex4}

\usepackage{graphicx}
\usepackage{dcolumn}
\usepackage{bm}

\begin{document}


\title{Doping dependence of chemical potential and entropy in hole- and electron-doped high-$T_c$ cuprates}

\author{T. Tohyama}
\affiliation{Institute for Materials Research, Tohoku University, Sendai 980-8577,
Japan}
\email{tohyama@imr.tohoku.ac.jp}
\homepage{http://www.maekawa-lab.imr.tohoku.ac.jp/}
\author{S. Maekawa}
\affiliation{Institute for Materials Research, Tohoku University, Sendai 980-8577,
Japan}

\date{\today}

\begin{abstract}
We examine the thermodynamic properties of the hole- and electron-doped cuprates by using the $t$-$t'$-$t''$-$J$ model.  We find that the chemical potential shows different doping dependence between the hole and electron dopings.  Recent experimental data of the chemical potential shift are reproduced except for lightly underdoped region in the hole doping where stripe and/or charge inhomogeneity are expected to be important. The entropy is also calculated as a function of the carrier concentration. It is found that the entropy of the electron-doped system is smaller than that of the hole-doped systems. This is related to a strong antiferromagnetic short-range correlation that survives in the electron-doped system.
\end{abstract}

\pacs{74.25.Bt, 71.10.Fd, 74.72.Dn}
\maketitle

High-$T_c$ superconductivity emerges with carrier doping into insulating cuprates.  The carrier is either an electron or a hole.  The phase diagram exhibits an asymmetry between the electron and hole dopings: in the electron-doped cuprate Nd$_{2-x}$Ce$_x$CuO$_4$ (NCCO), an antiferromagnetic (AF) phase remains up to the concentration $x\sim0.15$, while in the hole-doped cuprate La$_{2-x}$Sr$_2$CuO$_4$ (LSCO) the AF phase disappears with an extremely small amount of $x$.~\cite{Takagi}  Remarkable differences of the electronic states between the two materials have been observed in several experiments.  Inelastic neutron scattering experiments showed the presence of incommensurate spin structures in LSCO but not in NCCO.~\cite{Yamada}  The optical conductivity exhibits a gaplike feature at around 0.2~eV in the AF phase of NCCO,~\cite{Onose} but there is no such feature in LSCO.  From angle-resolved photoemission experiments, it is clearly observed that hole carriers doped into the parent Mott insulators first enter into the ($\pm\pi/2$,$\pm\pi/2$) points in the Brillouin zone, but electron carriers are accommodated at ($\pm\pi$,0) and (0,$\pm\pi$).~\cite{Armitage}  The doping dependence of the core-level photoemission also shows different behaviors of the chemical potential shift between NCCO and LSCO.~\cite{Harima1}  It is interesting that even in hole-doped cuprates the chemical potential shift strongly depends on materials~\cite{Harima2}: the shift is larger in Bi$_2$Sr$_2$Ca$_{1-x}$(Pr,Er)$_{x}$Cu$_2$O$_{8+y}$ (BSCCO) than in LSCO.  This indicates the difference of the electronic states among the hole-doped cuprates.

In previous studies,~\cite{Tohyama1} we showed that the $t$-$J$ model with long-range hoppings $t'$ and $t''$ explains the differences of the inelastic neutron scattering, optical conductivity, and angle-resolved photoemission data between hole- and electron-doped cuprates.  In this paper, we examine the thermodynamic properties of the hole- and electron-doped cuprates by using the same model.  A finite-temperature version of the Lanczos method for small clusters is employed to calculate the thermodynamic properties under the grand canonical ensemble. The calculated chemical potential shows a different dependence on the carrier concentration between the hole and electron dopings.  The experimental data~\cite{Harima1,Harima2} are nicely reproduced except for a lightly underdoped region in the hole doping where stripe and/or charge inhomogeneities are expected to play important roles.  The entropy is also calculated as a function of the carrier concentration. It is found that the entropy in NCCO is smaller than the entropies of LSCO and BSCCO.  A strong AF short-range correlation that survives in the electron-doped system is the origin of the small entropy.

The $t$-$J$ Hamiltonian with long-range hoppings, termed
the $t$-$t'$-$t''$-$J$ model, is
\begin{eqnarray}
H&=& J\sum\limits_{\left<i,j\right>_{1{\rm st}}}
      {{\bf S}_i}\cdot {\bf S}_j
    -t\sum\limits_{\left<i,j\right>_{1{\rm st}} \sigma }
    c_{i\sigma }^\dagger c_{j\sigma } \nonumber \\
&& {} -t'\sum\limits_{\left<i,j\right>_{2{\rm nd}} \sigma }
    c_{i\sigma }^\dagger c_{j\sigma }
     -t''\sum\limits_{\left<i,j\right>_{3{\rm rd}} \sigma }
    c_{i\sigma }^\dagger c_{j\sigma }+{\rm H.c.}\;,
\label{H}
\end{eqnarray}
where the summations $\left< i,j \right>_{1{\rm st}}$, $\left< i,j \right>_{2{\rm nd}}$ and $\left< i,j \right>_{3{\rm rd}}$ run over first-, second- and third-nearest-neighbor pairs, respectively.  No double occupancy is allowed, and the rest of the notation is standard.  In the model, the difference between hole and electron carriers is expressed as the sign difference of the hopping parameters~\cite{Tohyama2} $t>0$, $t'<0$, and $t''>0$ for hole doping, and $t<0$, $t'>0$, and $t''<0$ for electron doping.  The ratios $t'$/$t$ and $t''$/$t$ are taken to be material dependent: ($t'/t,t''/t$)=($-0.34,0.23$) for NCCO and BSCCO, and ($-0.12,0.08$) for LSCO.~\cite{Tohyama2}  $J$/$|t|$ is taken to be 0.4.  In order to examine the thermodynamic properties of the model, we use a finite-temperature version of the Lanczos method~\cite{Jaklic} for a square lattice with $\sqrt{18}$$\times$$\sqrt{18}$ sites under periodic boundary conditions.~\cite{Size}  The chemical potential $\mu$ and the entropy density $s$ are calculated under the grand-canonical ensemble.

\begin{figure}
\begin{center}
\includegraphics[width=8.cm]{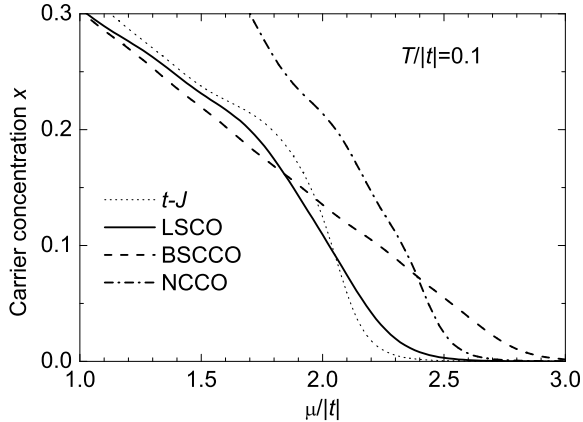}
\caption{\label{fig1}
Carrier concentration $x$ vs chemical potential $\mu$ for several parameter sets of the $t$-$t'$-$t''$-$J$ model with 18 sites.  $T/\left| t\right|=0.1$ and $J/\left| t\right|=0.4$.  The hopping parameters are ($t,t',t''$)=($1,0,0$) for $t$-$J$ (dotted line), ($1,-0.12,0.08$) for LSCO (solid line), ($1,-0.34,0.23$) for BSCCO (dashed line), and ($-1,0.34,-0.23$) for NCCO (dot-dashed line).}
\end{center}
\end{figure}

Figure~1 shows the carrier concentration $x$ for the $t$-$t'$-$t''$-$J$ model with different parameter values as a function of $\mu$ at $T=J/4=0.1\left| t\right|$.  The data for the $t$-$J$ model are consistent with previous reports.~\cite{Jaklic}  With increasing magnitudes of $t'$ and $t''$ in the hole carrier side (from $t$-$J$ to LSCO and BSCCO), the slope of the $x$ vs $\mu$ curves at a small concentration region $x<0.2$ becomes weak.  The derivative of $x$ with respect to $\mu$ is proportional to the charge compressibility $\kappa\propto -{\partial x}/{\partial \mu}$.  The variation of the slope thus means that $\kappa$ decreases with increasing $t'$ and $t''$, i.e., in the order $t$-$J$, LSCO, and BSCCO.  The fact that the charge fluctuation weakens with the increase of long-range hoppings is consistent with the tendency that the phase-separated region in the $x$ vs $J/t$ phase diagram  at zero temperature is reduced with increasing $t'$.~\cite{Tohyama3}  On the other hand, the slope in NCCO is similar to that of $t$-$J$, indicating that the charge fluctuation is as strong as that in $t$-$J$.  Comparing NCCO and BSCCO, both of which have the same magnitude of $\left| t'\right|$ and $\left| t''\right|$, we can see a remarkable difference in the doping dependence of $\mu$ between the electron- and hole-doped systems.

\begin{figure}
\begin{center}
\includegraphics[width=8.cm]{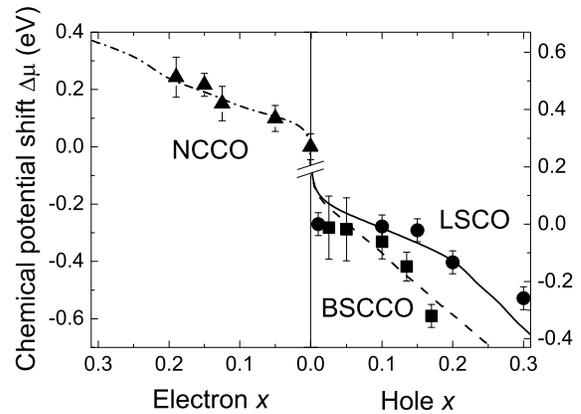}
\caption{\label{fig2}
Chemical potential shift $\Delta\mu$ vs carrier concentration $x$ for both hole- and electron-doped systems.  The solid, dashed, and dot-dashed lines represent the calculated results for the parameters of LSCO, BSCCO, and NCCO, respectively.  The value of $\left| t\right|$ is set to be 0.35~eV for all systems.  $J/\left| t\right|=0.4$ and $T/\left| t\right|=0.1$.  The circles and triangles denote the experimental data of LSCO and NCCO taken from Ref.~5, respectively, and the squares are the BSCCO data from Ref.~6.  The experimental shift is measured from $\mu$ at the lowest concentration examined in the experiments.}
\end{center}
\end{figure}

In Fig.~2, the experimental data of the chemical potential shift $\Delta\mu$ (Refs.~5 and 6) are replotted for the sake of a comparison with our theoretical results.  The experimental shift is measured from $\mu$ at the lowest concentration in each panel.  In the lightly doped region, almost no change of $\Delta\mu$ is observed in the hole-doped materials, while in NCCO $\Delta\mu$ is proportional to $x$.  It is also interesting in the experimental data that the two hole-doped materials LSCO and BSCCO exhibit different behaviors: the change of $\Delta\mu$ at around $x\sim0.1$ is larger in BSCCO than in LSCO.

In order to compare our results in Fig.~1 with the experimental $\Delta\mu$, we use the following procedures:  (i) We assume that $\left| t\right|$ is material independent and has $\left| t\right|=0.35$~eV.~\cite{Tohyama2} (ii) We examine a carrier concentration $x^*$ where $\mu$ is almost temperature independent,~\cite{Jaklic} and fit $\mu$ at $x^*$ to an experimentally expected $\Delta\mu$ at $x^*$: ($x^*,\Delta\mu$)=($0.18,-0.1$~eV), ($0.24,-0.4$~eV), and ($0.16,0.2$~eV) for LSCO, BSCCO, and NCCO, respectively.  We find in Fig.~2 that our results nicely reproduce the global features of the experimental $\Delta\mu$ in both electron- and hole-doped materials.  In particular, the calculated data show a good agreement with the experimental data at around $x\sim$0.1 in both LSCO and BSCCO.  Therefore, the different chemical potential shifts between the two materials can be attributed to the difference in the long-rang hoppings.  Although the global agreement with experiment is satisfactory, we find remarkable deviations from the experimental data in the lightly underdoped regions of the hole-doped systems ($x<0.1$), where experimentally ${\partial \mu}/{\partial x}\sim 0$ ($\kappa\rightarrow\infty$).  One of the possible origins of the deviations might be the difference of the temperature between the measurements ($T\sim80$~K) and the calculations ($T=0.1\left| t\right|=350$~meV $\sim400$~K).  If we were able to reduce $T$ by an increase in the system size, the deviations would become small because of the enhancement of $\kappa$.~\cite{Jaklic}  However, it is unlikely that $\kappa$ diverges independently of the magnitude of $t'$ and $t''$.  In this context, we hope for experiments at higher temperatures.  Another possibility that can account for the deviations might be the presences of stripes and/or charge inhomogeneity that are experimentally reported.~\cite{Singer,Pan}  Since we have no evidence of the charge inhomogeneity in our calculations, it may be necessary to calculate $\mu$ in the presence of external potentials that induce such an inhomogeneity.  This would be a future problem to be solved.

\begin{figure}
\begin{center}
\includegraphics[width=8.cm]{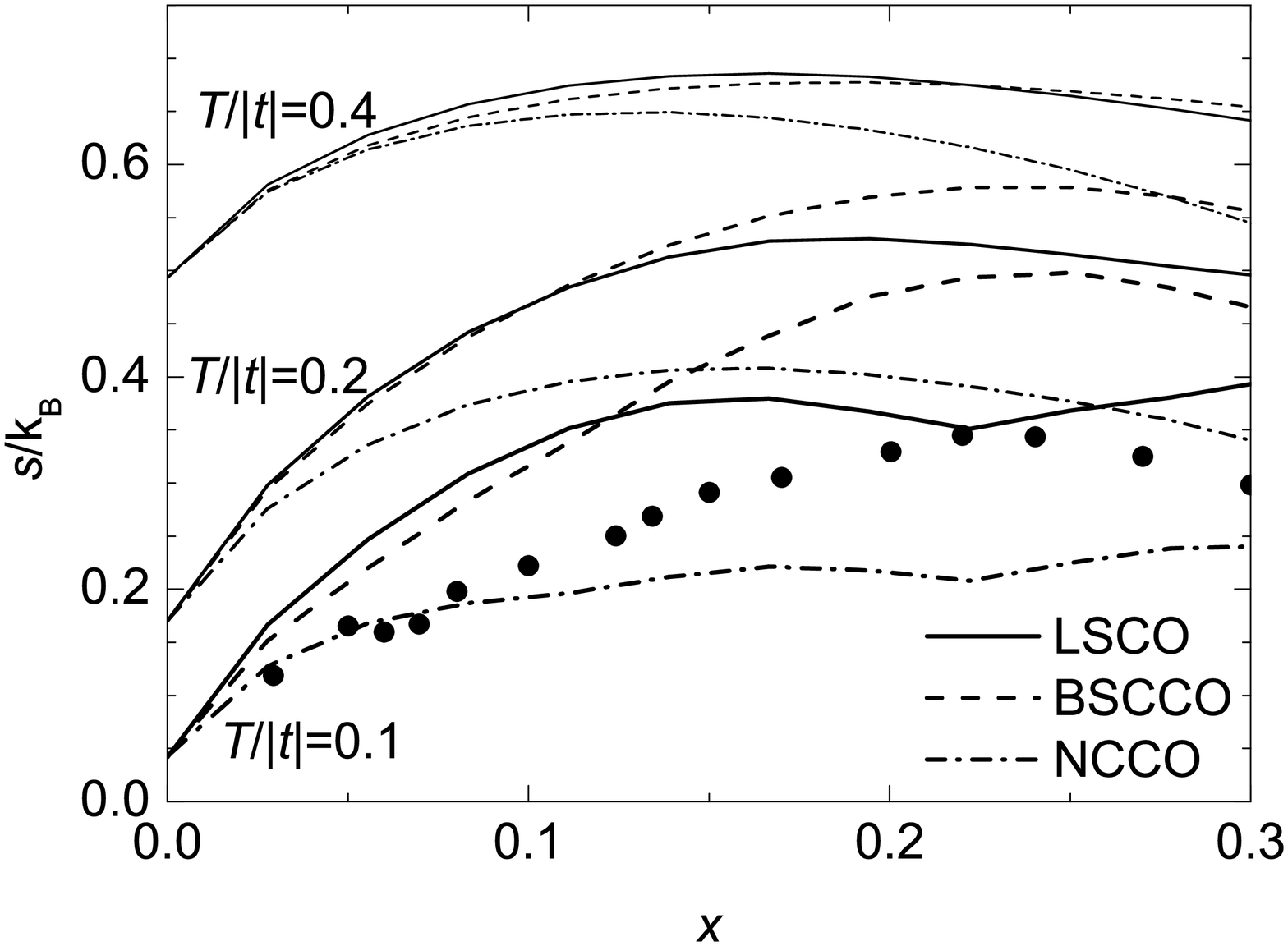}
\caption{\label{fig3}
Entropy density $s$ vs carrier concentration $x$ for both hole- and electron-doped systems with 18 sites.  The solid, dashed, and dot-dashed lines represent the results for the parameters of LSCO, BSCCO, and NCCO, respectively. $J/\left| t\right|=0.4$.  The three sets of the three lines correspond to the data at $T/\left| t\right|$=0.1, 0.2, and 0.4 from the bottom to the top.  The circles shows experimental results for LSCO at $T=320$~K, taken from Ref.~14.}
\end{center}
\end{figure}

Figure~3 shows the doping dependence of the entropy density $s$ at various temperatures for the three parameter sets of the $t$-$t'$-$t''$-$J$ model.  At $T/\left| t\right|=0.4$, there is no remarkable difference among LSCO, BSCCO, and NCCO, particularly in the underdoped region.  With decreasing temperature from $T/\left| t\right|=0.4$ to 0.1, $s$ in NCCO is strongly suppressed as compared with LSCO and BSCCO.  We find by examining the temperature dependence of the spin correlation~\cite{Tohyama1} that the suppression of $s$ in the underdoped region is correlated with the development of the AF short-range order.  This is easily understood because the AF order reduces the entropy coming from the spin degree of freedom.  On the other hand, the difference of $s$ between hole- and electron-doped systems in the overdoped region is probably due to the difference of the density of states.  It is desirable to confirm the suppression of $s$ in electron-doped systems experimentally.  In Fig.~3, we also plot $s$ measured for LSCO at $T=320$~K.~\cite{Loram}  The agreement with the calculated LSCO data at $T/\left| t\right|=0.1$ is qualitatively good, but not quantitatively satisfactory.  There are several reasons for the disagreement: (i) uncertainties in the conversion of a theoretical $T$ into a realistic one and in the experimental determination of $s$, (ii) a finite-size effect in our calculations that is seen as the dip of $s$ at $x\sim 0.22$, which comes from relatively large sparseness of the low-energy levels in the 18-site four-hole system, and (iii) more plausibly the effect of the stripe and/or charge inhomogeneity discussed above, by which $s$ in the underdoped region is expected to be reduced.~\cite{Tohyama4}

In summary, we have examined the thermodynamic properties of the hole- and electron-doped cuprates by using the $t$-$t'$-$t''$-$J$ model.  The calculated chemical potential shows different behaviors between the hole and electron dopings.  The experimental data of the chemical potential shift are explained by taking into account the material dependences of $t'$ and $t''$, except for the lightly underdoped region in the hole doping where the stripe and/or charge inhomogeneities are expected to be important. The entropy is also calculated as a function of the carrier concentration. It is found that the entropy of the electron-doped system is smaller than that of the hole-doped ones. This is related to a strong AF short-range correlation that survives in the electron-doped system.  To confirm this, specific heat measurements in the electron-doped materials are desired.

We would like to thank N. Harima and A. Fujimori for sending us experimental data prior publication and useful discussions.  This work was supported by a Grant-in-Aid for Scientific Research from the Ministry of Education, Culture, Sports, Science and Technology of Japan, and CREST.  The numerical calculations were performed in the supercomputing facilities in ISSP, University of Tokyo, IMR and Information Synergy Center, Tohoku University.


\begin{thebibliography}{99}
\bibitem{Size}We have also examined a 4$\times$4-site lattice and obtained results similar to those mentioned in the text.
\end{thebibliography}
\end{document}